\documentstyle[prl,aps,epsf,twocolumn]{revtex}

\begin{document} 
\bibliographystyle{prsty} 
\draft
\tightenlines

\title{Winding number instability in the phase-turbulence 
regime of the Complex 
Ginzburg-Landau Equation } 

\author{ R. Montagne\cite{vagaraul}, 
E. Hern\'andez-Garc\'\i a 
, and  M. San Miguel }     
\address{Departament de F\'{\i}sica, 
Universitat de les Illes Balears,  \\
and Instituto Mediterraneo de Estudios Avanzados, IMEDEA (CSIC-UIB)\\ 
E-07071 Palma de Mallorca (Spain) \\
URL: http://formentor.uib.es/Nonlinear/
} 

\date{\today} 

\maketitle

\begin{abstract} 
We give a statistical characterization of states with nonzero winding number
in the Phase Turbulence (PT) regime of the one-dimensional Complex 
Ginzburg-Landau equation. We find that states
with winding number larger than a critical one are unstable, in the sense that
they decay to states with smaller winding number. The transition
from Phase to Defect Turbulence is interpreted as an ergodicity breaking transition which occurs when  the 
range of stable winding numbers vanishes. Asymptotically stable states
which are not spatio-temporally chaotic are described
within the PT regime of nonzero winding number.
\end{abstract} 

\vskip 0.4cm


Spatio-temporal complex dynamics \cite{CrossHohenberg,CrossHohenberg2} is 
one of the present focus 
of research in nonlinear phenomena. Much effort has been 
devoted to the characterization 
of different dynamical phases and transitions between them for model 
equations such as the Complex Ginzburg-Landau Equation 
(CGLE)
\cite{CrossHohenberg,chate1,chate2,chate3,janiaud1,hohenbergsaarloos,egolf194,chate8,egolf195,montagne196}. 
One of the main 
questions driving these studies is whether concepts brought from statistical
mechanics can be useful for describing complex nonequilibrium systems
\cite{chate1,hohenberg89}. In this paper we give a 
characterization of the spatio-temporal configurations that occur in the 
Phase Turbulence (PT)
regime of the CGLE (described below), for a finite system, in terms of a global
wavenumber.  This quantity plays the role of an order parameter classifying
different phases. We show that in the PT regime there is an instability
such that a conservation law for the
global wavenumber occurs only for wavenumbers within a finite range that
depends on the point in parameter space. Our study is statistical in the sense 
that averages over 
ensembles of initial conditions are used. Our results allow a characterization 
of the transition from PT  to Defect or Amplitude Turbulence (DT) 
(another known dynamical regime
of the CGLE) as the line in parameter space in which the range of conserved 
global wavenumbers shrinks to zero.

The CGLE is an amplitude 
equation for a complex field $A({\bf x},t)$ 
describing universal features of the dynamics of extended systems near a Hopf
bifurcation \cite{CrossHohenberg,hohenbergsaarloos}.   
\begin{equation} 
 \partial_{t} A =  A + (1 + {\it i} c_1 ) \nabla^{2} A -  
 (1 + {\it i} c_2 ) \mid A \mid^{2} A \ . 
 \label{cgle}  
\end{equation}  
Binary fluid convection \cite{kolodner95}, transversally extended lasers
\cite{coullet}, chemical turbulence\cite{kuramoto81}, bluff body wakes 
\cite{provansal94}, among other systems can be described by the CGLE in the
appropriate parameter range. We will restrict ourselves in this paper to the
one-dimensional case, that is $A=A(x,t)$, with $x \in [0,L] $.  For this
situation a major step towards the analysis of phases and
phase transitions in (\ref{cgle}) was the identification 
\cite{chate1,chate2,chate3} of different chaotic regimes in
different regions of the parameter space $[c_1,c_2]$ (see Figure 1).
\begin{figure} 
\vskip 5.0 cm 
\includegraphics{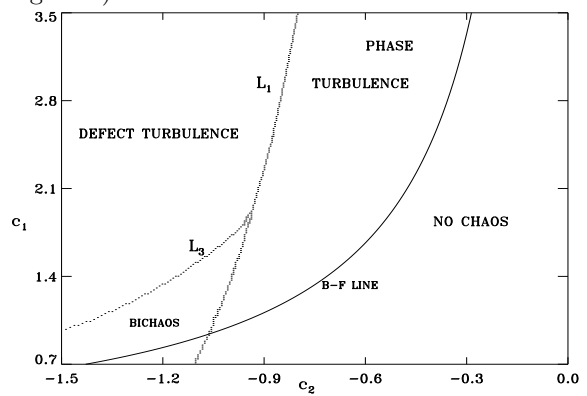} 
\caption{Regions of the parameter $[c_{1}, c_{2}]$-space for the
CGLE displaying different kinds of regular and  chaotic behavior. Lines
$L_1, L_3$ were determined in\protect\cite{chate1,chate2,chate3}. }
\end{figure}
Eq. (\ref{cgle}) has plane-wave solutions $A_k=\sqrt{1-k^2}e^{ikx}$ with
$k \in [-1,1]$. When $c_1c_2>-1$ there is a range of wavenumbers 
$[-k_E, k_E]$ such that the
plane-wave solutions with wavenumber in this range are linearly stable. 
They become unstable outside this range (the Eckhaus instability
\cite{janiaud1}). The limit of this range, $k_E$ approaches zero as
the product $c_1c_2$ approaches $-1$, so that the range of stable plane  waves
vanishes by approaching from below the line $ c_1c_2=-1$ (the Benjamin-Feir 
or Newell line, labeled BF in Fig.1). Above that line no plane wave is stable 
and different turbulent states exist. The authors of 
\cite{chate1,chate2,chate3} identified three different regimes in different 
regions above the BF line (Fig.1): PT, DT, and  bichaos. 
Among these regimes, the transition between PT and DT has received special 
attention \cite{chate1,egolf195,sakaguchi2}. 
In spite of the fact  that there
are some indications that this transition can be ill-defined in the
$L\rightarrow\infty$ limit \cite{chate3,chate8,egolf195}, the PT regime is
robustly observed for any finite size system and for finite observation times, 
with the transition to DT appearing at a quite well defined line 
($L_1$ in Fig.1) \cite{chate8}. 
In the
DT region the modulus $|A|$ of $A=|A|e^{i\phi}$ becomes zero at
some  instants and places (called {\sl defects}), so that the phase $\phi$
becomes undefined and the winding number $\nu \equiv \frac{1}{2 \pi} \int_0^L
\partial_x \phi dx$ changes  value during evolution. 
In contrast, dynamics
maintains the modulus of $A$ far from zero in the PT region, so 
that $\nu$ is thought to be a constant of motion there. A global wave number of
the configuration can be defined as $k\equiv 2\pi \nu/L$. These different 
regimes were 
originally identified from the analysis of the space-time density of  
{\sl defects}. If this picture is correct, one can speculate that the 
transition 
between DT and PT would be a kind of ergodicity breaking transition 
\cite{palmer89} as in other systems described by statistical mechanics. DT
would correspond to a ``disordered" phase and  
$\nu$  classifies different ``ordered" phases in PT. However, we note that 
most studies of
the PT regime have only considered in detail the case of $\nu=0$. In
fact the phase diagram in Fig.1 was constructed for this case.
In order to provide a better understanding of the PT-DT
transition we undertake in this Letter a systematic study of PT configurations 
with $\nu \not = 0$ .

Typical configurations of the PT state of zero
winding number consist of
pulses in $|A|$, corresponding to phase sinks, that travel and collide rather
irregularly on top of a $k=0$ unstable background wave (that is, a 
uniform oscillation)\cite{chate1,chate3}. 
The phase of these configurations strongly resembles solutions  of  the 
Kuramoto-Shivashinsky (KS) equation. Quantitative 
agreement has been found between the $\nu=0$ PT states of the CGLE 
and solutions of the  KS equation near the BF line\cite{egolf195}.  
 The more obvious 
effect of a non-zero
$\nu$ is the appearance of a uniform drift added to the irregular motion of 
the pulses. In addition 
Chat\'e\cite{chate2,chate3} reported an earlier breakdown of the PT regime
when $\nu \not = 0$. Our results below show
that not all the  winding numbers are in fact allowed in the PT region at
long times.
PT states with too large $|\nu|$ are only transients and decay to states 
within a
band  of allowed winding numbers. The width of this band shrinks to zero when
approaching the line $L_1$. In addition we find that the allowed non-zero 
winding number states are not of  a single type. We have identified three 
basic types of asymptotic states for $\nu \neq 0$. Two of them are different 
from the usually described
PT states in the sense that they do not exhibit spatio-temporally chaos.

In order to study the dynamics of states with  $\nu \neq 0$ we have performed 
simulations extensively covering 
the PT region of parameters of Fig. 1. Only a small part of the simulations 
is shown here, and the rest will be reported elsewhere. We use a 
pseudo-spectral code with
periodic boundary conditions and second-order accuracy in time. Spatial
resolution was typically 512 modes, with runs of up to 4096 modes to
confirm the results. We work at fixed system size $L=512$.  
The initial condition in our simulation is a plane wave of the desired winding 
number, slightly perturbed by a white Gaussian random field. The initial 
evolution of
the spatial power spectrum is well described by the linear stability analysis 
around the initial plane wave: typically
the perturbation grows mostly around the most unstable wavenumbers identified
from such linear analysis. After some time the
system reaches a state similar to the $\nu=0$ PT, except for 
a non-zero average velocity of the chaotically travelling pulses. We call 
this state {\sl riding PT}. Its spatial power spectrum is broad and unsteady, 
with the more active wavenumbers located
around the one determined by the initial winding number. We observe that when 
this winding number
is small, it remains constant in time, and the system either remains in the
{\sl riding PT} state or approaches one of the more regular 
asymptotic states that will be described below. If $|\nu|$ is initially  
 too high, the competition between wavenumbers leads to phase slips that 
reduce $|\nu|$ until a value inside an allowed range is reached. Then the system evolves
as before.  

We present in Fig. 2 the temporal evolution of $\bar\nu(t)$, the average of 
$\nu(t)$ over 50 independent realizations of the random perturbation added to 
the initial plane wave for a fixed point in parameter space. 
The variance among the sample of 50 realizations is also shown. Three initial 
values $\nu_i$ of the winding number are shown.
$\bar\nu(t)$ typically presents a decay  from $\nu_i$ to the final winding
number $\nu_f$. 
\begin{figure} 
\vskip 5.3 cm 
\includegraphics{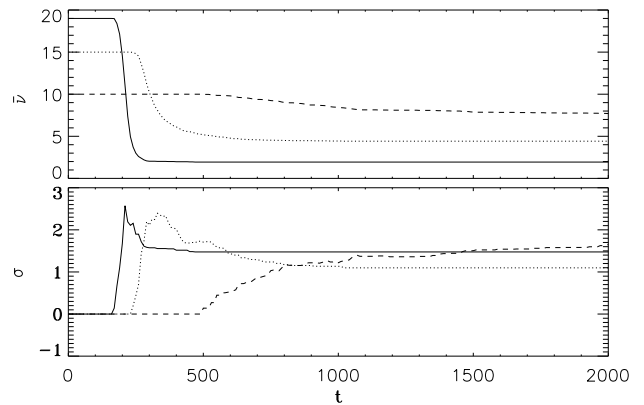}
\caption{a) Temporal evolution  of $\bar\nu (t)$ for three different initial
winding numbers $\nu_i = 19$ (solid), $15$(dotted), $10$ (dashed) . 
$c_{1} = 2.1, 
c_{2} = -0.75$. b) Winding number standard deviation $\sigma$.}
\end{figure}
The decay is found to take place in a characteristic time $\tau$ that we 
quantify as the time for which half of the jump in $\nu$ has been attained. 
Figure 3 shows $1/\tau$ for different values of $\nu_i$. The different curves
correspond to different values of $c_2$ with fixed $c_1$.
Similar results were obtained for $c_2$  fixed and varying 
$c_1$. $\tau$ 
increases with an apparent divergence as $\nu_i$ 
approaches a particular value $\nu_c$ which is a function of $c_1$ and $c_2$. 
We estimate
this $\nu_c$ by fitting linearly the data for $1/\tau$. Other fits involving
non-trivial critical exponents have been tried, but they do not improve 
the simpler linear one in a significant manner. A very similar value of 
$\nu_c$ is obtained by simply determining the  value of
$\nu_i$ below which $\nu(t)$ does not change in any of the realizations.
Values of $\nu_c$ from some of the simulations are in the inset of Fig. 3.  
$\nu_c$ vanishes as $c_2$ approaches  the transition line $L_1$ (or $L_3$ when
passing through the bichaos region). For example, 
the linear fitting of the data in the inset of Fig.3 and extrapolation 
towards zero $\nu_c$ reproduces the value for $L_1$ of 
\cite{chate1,chate3} 
($c_2 \approx -0.9$ for $c_1= 2.1$) within the fitting error in 
$c_2$ of $\pm 0.02$. 

\begin{figure} 
\vskip 5.5 cm 
\includegraphics{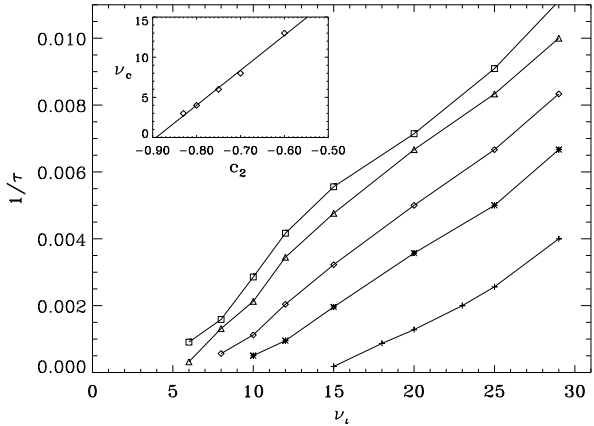}
\caption{Inverse of the characteristic time for winding number relaxation 
as a function of the initial 
winding number. The value of $c_{1}$ is fixed ($c_1=2.1$) and  $
c_{2}$ varies from near the B-F line ($c_2 = -1/2.1$) to the $L_1$ line 
($c_2 \approx -0.9$). Different symbols correspond to $c_2= -0.6$ ($+$), 
$c_2 =-0.7$ ($\ast$), 
$c_2 = -0.75$ ($\Diamond$), $c_2 = -0.8$ ($\triangle$), $c_2= - 0.83$
 ($\Box$). The inset 
shows the critical winding number  ($\nu_c$) as a function of  $ c_{2}$.  
}
\end{figure}

The winding number instability found here in the PT region is strikingly 
similar to  the Eckhaus
instability of travelling waves below the BF line of  Fig. 1
\cite{janiaud1}: There is a range of allowed
winding numbers such that configurations outside this range undergo phase slips until 
an allowed $\nu$ is reached. The difference is that below the BF line, the
attractor for each stable $\nu$ is a travelling plane wave of wavenumber
$k$, whereas each $\nu$, or equivalent global wave number, characterizes  
phase turbulent attractors above the BF line. The allowed range of 
travelling waves shrinks to zero 
when $(c_1,c_2)$ approaches the BF from below, whereas above BF, the allowed 
$\nu$ range shrinks to zero when approaching the $L_1$ line from the right. 
In this picture, the transition PT-DT appears as the {\sl BF line} associated 
to an Eckhaus-like instability for phase turbulent waves. Since the
presence of chaotic fluctuations is a kind of self generated ``noise" present 
in the system,
this winding number instability should be compared to the Eckhaus instability
in the presence of stochastic noise \cite{emilio93},
rather than to the deterministic one. The comparison shows
qualitative similarities between both cases, but quantitative agreement 
such as similar critical exponents or scaling laws has not been found. 
The comparison is also instructive because it can be shown that, for the
one-dimensional stochastic case, there 
is no true long range order, and therefore no true phase transition in the 
infinite size limit \cite{emilio91}. But for finite sizes and finite
observation times well defined effective transitions and even critical
exponents can be introduced\cite{emilio93}. The PT-DT transition in the CGLE 
can be an effective transition of this kind. In order to further characterize
the robustness of the effective transition an analysis of system size effects 
should be performed.  Preliminary results 
indicate that the $\nu_c$ obtained  for 
each $(c_1,c_2)$ point grows linearly with system size $L$, as it should happen
for a well defined extensive quantity.  

Finally we consider the nature of the asymptotic states allowed within 
the band of ``stable" $\nu$. \\

 \begin{figure} 
\vskip 10 cm 
\includegraphics{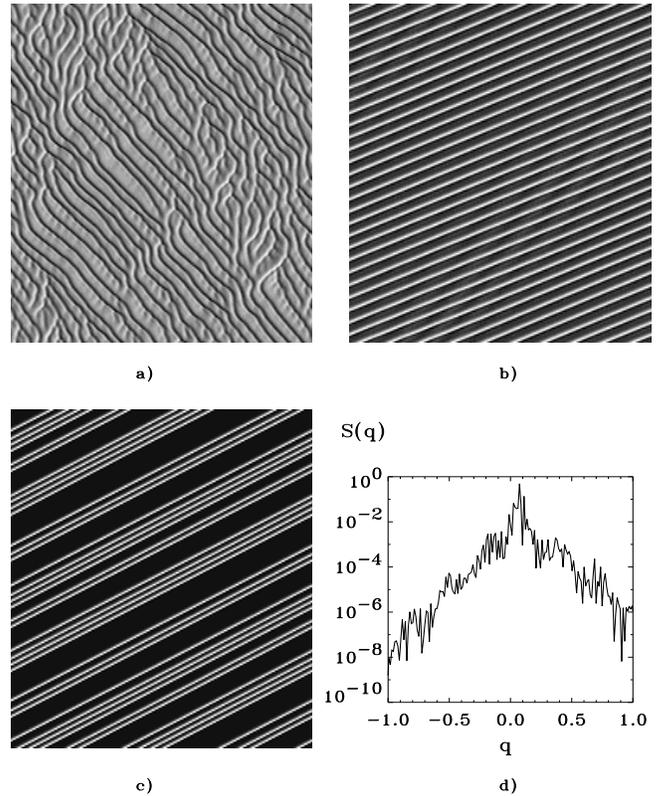}
\caption{Spatiotemporal evolution of $\partial_x\phi(x,t)$ with time running
upwards and $x$ in the horizontal direction.  The lighter
grey correspond to the maximum value of $\partial_x\phi(x,t)$ and darker to 
the minimum. Different scales
of grey are used in each case in order to see the significant structures.
a) Last  $10^2$ time units of a run $10^4$ time units long  
for a {\sl riding PT} state at $c_1 = 2.1$ , and $c_2 = - 0.83$. 
The  initial condition was a TW with  $\nu_i = 20$ that decayed to 
$\nu_f = -1$ after a short time.
b) Last $10^2$ time units of a run $10^5$  time units long for a 
quasiperiodic state. The
initial condition is random noise with an amplitude of $0.05$. 
$c_1 = 2.0$ , and $c_2 = -0.8$. 
c)Last $10^2$ time units of a run $10^4$ time units long for a frozen 
turbulence  state. The initial condition is a TW 
of $\nu_i = 12$ that decayed to $\nu_f = 6$ after a short time. $c_1 = 1.75$ , 
and $c_2 = -0.8$. 
d) Spatial power spectrum $S(q)$ as a function of wavenumber 
for the frozen turbulence configuration shown in c). This spectrum is constant in
time.}  
\end{figure}

 We have numerically found three basic types of 
states in the PT
region of parameters with non-zero $\nu$. Fig. 4 shows in gray levels the 
value of $\partial_x\phi(x,t)$ as a function of $x$ and $t$. The state 
shown in the top left is the
familiar\cite{chate3} {\sl riding PT}, which is similar to the PT usually seen
for $\nu=0$ (wiggling pulses in the gradient of the phase) except for a 
systematic drift in a direction determined by $\nu$. The other two states do not show spatio-temporal chaos. They can be described as the motion in time of a spatially 
rigid pattern on the top of
a plane wave (with $k \neq 0$) background and with periodic boundary 
conditions. The state shown in the top right consists of equidistant pulses
travelling uniformly.  They are the
quasiperiodic states described in \cite{janiaud1}.  The state shown in 
the bottom left, that we call {\sl frozen turbulence}
consists of pulses uniformly travelling  on a plane wave background, as in the
quasiperiodic case, but now the pulses are not equidistant from each  other.
The spatial power spectrum is shown for this later case.  It is a broad
spectrum in the sense that the inverse of its width, which gives a measure of
the correlation length, is small compared with the system size. This is due 
to the irregular positions of the pulses. In addition the spectrum is 
constant in time,
which makes this frozen state different from riding PT.  The existence 
of the two states with no
spatio-temporal chaos (quasiperiodic and frozen turbulence) 
described 
above can be understood by analyzing the phase equation valid near the BF
instability. In the case of a non-zero $\nu$ it contains terms breaking the
left-right symmetry \cite{janiaud1,sakaguchi1}, and it is known as Kawahara
equation \cite{kawa83}. Its uniformly travelling solutions are related to the
rigidly propagating patterns of Fig.~1b) and c). These solutions can be
analyzed with the tools of Shilnikov theory \cite{wiggins1}. The details 
will be discussed elsewhere.  

In addition to the pure three basic states there are configurations in which 
they coexist at different places of space, giving rise to a kind 
of {\sl intermittent} configurations, some of them already observed in
\cite{chate2}. 
The main results reported here, that is the existence of an Eckhaus-like
instability for phase-turbulent waves, the identification of the transition
PT-DT with the vanishing of the range of stable winding numbers, and the
coexistence of different kinds of PT attractors should in principle be observed
in systems for which PT and DT regimes above a Hopf bifurcation are known to
exist \cite{provansal94}.  We note in addition that the experimental
observation of what seems to be an Eckhaus instability for non-regular waves has 
been already reported in \cite{debruyn94}. 

Financial support from DGYCIT (Spain) Projects PB94-1167 and PB94-1172 is
acknowledged. R.M. also acknowledges partial
support from the Programa de  Desarrollo de las Ciencias B\'asicas (PEDECIBA,
Uruguay), the Consejo Nacional de Investigaciones Cient\'\i ficas Y
T\'ecnicas (CONICYT, Uruguay) and the Programa de Cooperaci\'on con
Iberoam\'erica (ICI, Spain).


\end{document}